\documentclass[a4paper, 10pt, conference]{ieeeconf}      
\usepackage{FG2025}
\usepackage{subcaption}
\usepackage{graphicx}
\usepackage{multirow}
\usepackage{multicol}
\usepackage{pgfplots}
\usepackage{tikz}
\usepackage{calc}
\usetikzlibrary{fit,calc}
\usepackage{amsmath} 
\usepackage{booktabs} 
\usepackage{siunitx}
\usepackage{colortbl}
\usepackage{balance}
\usepackage{fancyhdr}
\usepackage{url}
\FGfinalcopy 

\definecolor{lightred}{RGB}{255, 204, 204}
\definecolor{lightblue}{RGB}{204, 229, 255}
\definecolor{lightgreen}{RGB}{204, 255, 204}
\definecolor{lightyellow}{RGB}{255, 255, 204}
\definecolor{lightgray}{gray}{0.9}

\IEEEoverridecommandlockouts                              
\def\FGPaperID{214} 

\title{Shielding Latent Face Representations From Privacy Attacks}

\author{\parbox{16cm}{\centering
    {\large Arjun Ramesh Kaushik$^1$, Bharat Chandra Yalavarthi$^1$,\\Arun Ross$^2$, Vishnu Boddeti$^2$, Nalini Ratha$^1$}\\
    {\normalsize
    $^1$ University at Buffalo, Buffalo, USA\\
    $^2$ Michigan State University, East Lansing, USA}}
    \thanks{This work was funded by the NSF Center For Identification Technology Research (CITeR).}
}

\begin{document}

\ifFGfinal
\thispagestyle{empty}
\pagestyle{empty}
\else
\author{Anonymous FG2025 submission\\ Paper ID \FGPaperID \\}
\pagestyle{plain}
\fi
\maketitle

\thispagestyle{fancy}

\begin{abstract}

In today’s data-driven analytics landscape, deep learning has become a powerful tool, with latent representations, known as embeddings, playing a central role in several applications. In the face analytics domain, such embeddings are commonly used for biometric recognition (e.g., face identification). However, these embeddings, or templates, can inadvertently expose sensitive attributes such as age, gender, and ethnicity. Leaking such information can compromise personal privacy and affect civil liberty and human rights. To address these concerns, we introduce a multi-layer protection framework for embeddings. It consists of a sequence of operations: (a) encrypting embeddings using Fully Homomorphic Encryption (FHE), and (b) hashing them using irreversible feature manifold hashing. Unlike conventional encryption methods, FHE enables computations directly on encrypted data, allowing downstream analytics while maintaining strong privacy guarantees. To reduce the overhead of encrypted processing, we employ embedding compression. Our proposed method shields latent representations of sensitive data from leaking private attributes (such as age and gender) while retaining essential functional capabilities (such as face identification). Extensive experiments on two datasets using two face encoders demonstrate that our approach outperforms several state-of-the-art privacy protection methods. 

\end{abstract}

\section{INTRODUCTION}

\label{sec:intro}

\begin{figure}[!ht]
    \centering
    \begin{subfigure}[b]{0.5\textwidth}
        \centering
        \includegraphics[width=\textwidth]{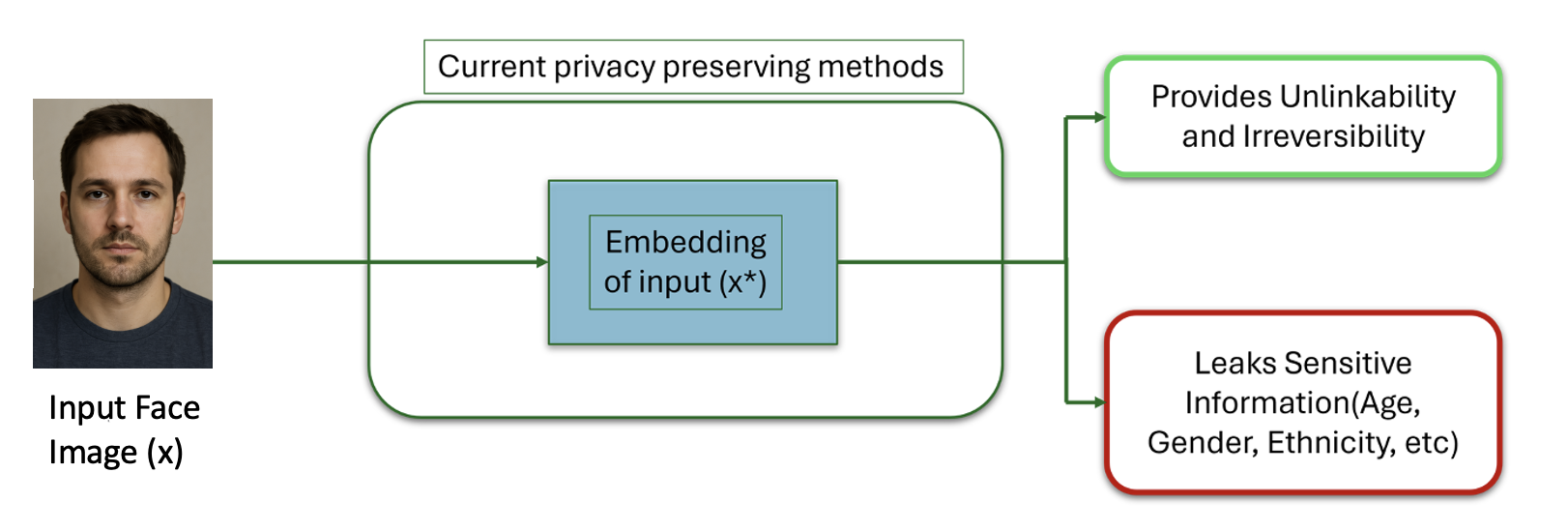}
        \caption{}
    \end{subfigure}%
    \hfill
    \begin{subfigure}[b]{0.5\textwidth}
        \centering
        \includegraphics[width=\textwidth]{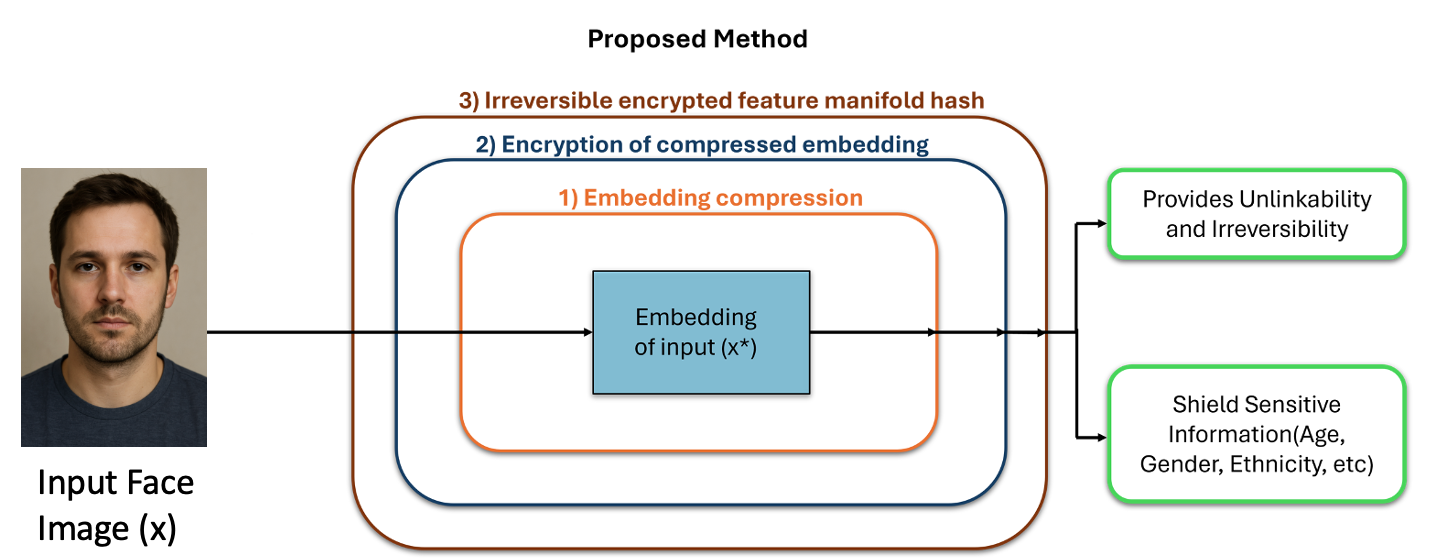}
        \caption{}
    \end{subfigure}
    \caption{(a) Existing embedding protection techniques typically guarantee unlinkability and irreversibility but leak sensitive information like age, gender, and ethnicity. (b) Our proposed method improves on the current methods by shielding the sensitive information against privacy attacks while still ensuring unlinkability and irreversibility. The face image used in the figure is AI-generated.}
    \label{fig:intro}
\end{figure}

The explosive advancement of deep neural networks (DNNs) has revolutionized computational capabilities across domains, from computer vision \cite{intro_img_analysis} and speech processing \cite{intro_speech_recog} to medical diagnostics \cite{intro_med_diagnosis}. Architectural innovations and optimization techniques have led to significant performance improvements. Research on the privacy and security of DNNs have gained momentum due to stringent regulatory frameworks such as the European Union's General Data Protection Regulation (GDPR) and US executive mandates \cite{ASISArticle}. These considerations are particularly crucial in domains that process sensitive information like biometric systems. Biometric systems use traits such as face, fingerprints, iris, or voice to automatically recognize individuals. In such fields, the trade-off between model utility and privacy guarantees demands precise quantification and optimization.

The core functionality of DNN models relies on projecting very high-dimensional input data into low-dimensional spaces, commonly known as latent representations or embeddings, to facilitate decision-making processes. Many critical applications, like biometric recognition, employ DNNs to improve recognition accuracy. Face embeddings have found widespread utility in facial analytics, spanning applications from law enforcement and border security to consumer device authentication \cite{fr_apps1}\cite{fr_apps2}\cite{fr_apps3}. While many of these applications serve beneficial purposes, the unrestricted deployment of these technologies without adequate privacy safeguards raises significant concerns. The potential for malicious exploitation of facial data presents tangible privacy risks to individuals \cite{EICHELBERG20201126}\cite{NASArticle}, emphasizing the need for privacy-preserving mechanisms in DNN models. 

Our research examines the privacy vulnerabilities of embeddings through three distinct attack vectors: Membership Inference, Attribute Inference, and Reconstruction Attacks. The seminal work on membership inference by Shokri et al. \cite{membership_inf} established a foundation for understanding data leakage in machine learning models. This led to subsequent investigations into reconstruction attacks, which aim to recover complete or partial training data \cite{recon1}\cite{recon2}. Parallel research has revealed the vulnerability of models to attribute inference attacks, where unintended characteristics can be extracted from the learned representations \cite{pu2024embeddingattackprojectwork}. For example, deep learning models trained for identity verification using face images can inadvertently expose user demographics like ethnicity, gender, and age. While acknowledging the broader landscape of adversarial machine learning, including adversarial examples and data poisoning, this work specifically addresses two fundamental privacy questions: the quantifiable extent of sensitive information extractable from embeddings, and the development of privacy-preserving mechanisms that maintain model utility.

To address the identified embedding vulnerabilities, we examine well-known privacy-preserving approaches. Differential Privacy techniques introduce controlled noise to combat property inference attacks \cite{differential_privacy_survey}, providing theoretical privacy guarantees. Another commonly employed technique is cancelable biometrics in authentication systems \cite{cb_intro}, where biometric data are encoded with an application-specific distortion scheme. Traditional encryption schemes effectively protect data at rest and during transmission \cite{9361564}\cite{Kiran_2020}, but they prohibit computational operations on encrypted data. Fully Homomorphic Encryption (FHE) \cite{attackOnFacial}\cite{FHEThreat} addresses this limitation by enabling direct computation on encrypted embeddings. 

\textbf{Contributions:} Our research systematically evaluates several privacy protection techniques, focusing specifically on Inversion Attacks (Membership Inference and Reconstruction) and Attribute Inference Attacks on embeddings. We conduct our analysis using a mix of face embedding models - AdaFace \cite{adaface} and ArcFace \cite{arcFace} -  while examining the leakage of soft biometric attributes (age, gender, ethnicity). Our key contributions are as follows:
(1) We establish that Differential Privacy with Laplacian noise successfully protects soft biometric attributes, but at the cost of degraded primary task performance in face recognition; (2) We examine how irreversible feature manifold hashes contribute to privacy in face embeddings. We explore Template Protection schemes as some irreversible feature manifold hashes. Although they effectively counter Inversion Attacks, they remain vulnerable to Attribute Inference Attacks. We validate these findings across three protection schemes available in the research literature: PolyProtect (PP), Negative Face Recognition (NFR), and Minimum Information Units (MIU) \cite{base} \cite{NFR}\cite{MIU}. (3) We introduce a novel multi-layer impenetrable shield (Fig. \ref{fig:intro} (b)) that achieves optimal privacy-utility balance. Our approach combines Fully Homomorphic Encryption (FHE) and an irreversible feature manifold hash on a compressed embedding to significantly enhance privacy while minimally impacting primary task performance.

\section{Threat Model}
\label{sec:threat_model}
We define \textit{image analytics} as the process of extracting semantic information from an image, which may include sensitive attributes such as age, gender, or ethnicity — commonly known as soft biometrics \cite{HealthCues2022Ross, medical2}. The potential to infer these soft biometric cues from facial images, or even from their embeddings, through automated techniques, poses privacy concerns~\cite{Dantcheva2016_what_else}. Such techniques, often implemented using machine learning models like SVMs or DNNs, can enable the extraction of personal information. For instance, if a biometric image (or its embedding) is compromised, it may allow unauthorized parties to derive sensitive biometric details, exposing private information of an individual.

While there are challenges in the training phase of the deep model, we focus on the leakage of side information during the inference stage. Consider a user who submits their face image or embedding to a cloud service to perform face recognition; the service provider can now learn several other details of the user beyond performing the main task. In our work, we presume that the embedding is provided in an encrypted form and the server has no access to the secret key to decrypt it. Our goal is to ensure that the encrypted embedding does not reveal any soft biometric information to unauthorized users. Note that the threat remains unchanged even if the parameters of the models used to extract soft biometric information (e.g., weights of a DNN) are encrypted.

We consider the most challenging threat model according to ISO/IEC 30316, which is the full disclosure model, where the attacker possesses complete knowledge of the irreversible feature manifold hash method for the embedding, including its algorithm, user-specific parameters, and one or more hashes corresponding to an embedding. In addition, we assume that the public key used in FHE is available, but not the private key. If the embeddings are {\em not} encrypted, the hacker can infer soft biometric attributes from the irreversible hash of the embeddings, as shown in Table \ref{TP_leakage}.

\subsection{FHE Basics}
\label{sec:fhe_basics}
Encryption is the process by which plain-text data is encrypted into ciphertext using a secret key and a cryptographic algorithm. Only authorized entities with a private key can decrypt the ciphertext back to the plaintext. Encryption is essential for securing sensitive data from unauthorized access or modification. Homomorphic encryption (HE) is a cryptographic system that permits certain computations to be performed on encrypted data without requiring decryption. In this system, we have public ($pk$) and private secret ($sk$) keys, encryption ($E$) and decryption ($D$) mechanisms, and plaintext values $x$ and $y$. When $x$ and $y$ are encrypted as $x' = E(x, pk)$ and $y' = E(y, pk)$, respectively, a cryptosystem is considered homomorphic with respect to a chosen operator (e.g., addition or multiplication), denoted as $◦$, if we can find another operator $•$ such that $x ◦ y = D(x' • y', sk)$. This means that we can conduct operations on encrypted data and obtain the same result when decrypting using the private secret key.

Specifically, given $c_i = \mathcal{E}(x_i, pk), i = 1, 2, \cdots, K$, an FHE scheme allows the computation of $c = g(c_1, c_2, \cdots, c_K)$ such that $\mathcal{D}(c, sk) = f(x_1, x_2, \cdots, x_K)$ for any arbitrary function $f$. 

It is essential to note that three types of homomorphic encryption exist \cite{he_taxonomy}: (i)
 Partial Homomorphic Encryption (PHE) permits addition or multiplication operations; (ii) Somewhat Homomorphic Encryption (SHE) allows limited computations on ciphertexts; (iii) Fully Homomorphic Encryption (FHE) enables computations on ciphertexts of any depth and complexity.

Numerous FHE systems have been introduced, among them are the ``Brakerski/Fan-Vercauteren" (BFV),  ``Brakerski, Gentry, Vaikuntanathan" (BGV), and ``Cheon, Kim, Kim and Song" (CKKS) schemes \cite{FHEACM}. The BFV and BGV schemes enable vector operations involving integers, while the CKKS scheme facilitates floating-point operations. These schemes achieve Single Instruction Multiple Data (SIMD) operations by bundling plaintext values into an array and then encrypting them to get ciphertext. In this work, we use the HEAAN \cite{HEAAN} library based on the CKKS scheme for FHE computations.

\section{Prior Work}
\label{sec:related_work}
We investigate a range of privacy-enhancing technologies in biometrics to safeguard sensitive user data.

\textbf{Privacy in Facial Analytics.} A significant body of research investigates methods for enhancing privacy of soft biometrics in face analytics at both the image and embedding levels. PFRNet \cite{PFRNet}, SensitiveNets \cite{morales2020sensitivenets}, PrivacyNet \cite{PrivacyNet}, and Multi-IVE \cite{mIVE}, each employ distinct approaches such as Autoencoder framework, adversarial regularization, and Incremental Variable Elimination to suppress sensitive attribute information in face embeddings. Unlike our proposed method, these approaches do not fully suppress soft biometrics leakage to the desired level of random guessing. Some work resort to disentangled representation learning as a privacy-preserving mechanism \cite{Jang_2024_CVPR}\cite{sarhan2020fairnesslearningorthogonaldisentangled}. However, disentanglement does not guarantee irreversibility and unlinkability.  

\textbf{Embedding Compression.} Compressing embeddings is another useful method for privacy enhancement, especially for deep embedding-based systems. Deep embeddings are often high-dimensional vectors that can contain privacy-sensitive information. Since compression can induce the loss of unwanted information, other sensitive information will likely become difficult to extract. Assine et al. \cite{assine2019compressing} propose a PCA-based compressing embedding representation. In \cite{eccv3}, a disentangling mechanism for graph embeddings is proposed to protect users' private information in social media data. \cite{Liao2020EmbeddingCW} present an adaptive bit allocation scheme which assigns different bits to dimensions in the embedding vector based on the importance of a dimension.

Based on their importance, Zhao et al. \cite{Liao2020EmbeddingCW} present an adaptive bit allocation scheme to assign different bits to different dimensions of embedding vectors. In our work, we use Matryoshka Representation Learning (MRL) \cite{matryoshka} optimized for face recognition to suppress sensitive attributes while maintaining recognition performance.

\textbf{Differential Privacy.} Differential privacy \cite{differential_privacy_survey} is often used for enhancing privacy in many application areas, including medical data \cite{DYDA2021100366}, social network analysis \cite{jiang2021applications}, online reviews \cite{eccv1}, and face recognition \cite{CHAMIKARA2020101951}\cite{Mao2018APD} \cite{face_diff_privacy_2023}. In \cite{eccv2}, a differential privacy mechanism is proposed to protect against inversion attacks from confidence scores. However, to our knowledge, no work has been reported that addresses the protection of ancillary data or the privacy of face embeddings.

\textbf{FHE for Facial Analytics and Medical Imaging.} Several papers \cite{reversing, attackOnFacial} have shown that existing soft biometric protection approaches are susceptible to reversibility and leakage. In our research, we demonstrate the efficacy of FHE in restricting soft biometric leakage in face embeddings to levels equal to or lower than random guessing and providing a more stringent theoretical guarantee of irreversibility compared to existing methods. 
FHE has been used in prior work with respect to face recognition. Several papers \cite{boddeti2018secure}\cite{pradel}\cite{ engelsma2022hers}\cite{ppFHE} propose various efficient approaches to perform face matching in the FHE domain using encrypted face embeddings for secure face authentication. In \cite{attackOnFacial}\cite{FHEThreat}, the authors show the susceptibility of homomorphic encryption to leaking soft biometric attributes when the private key is not secured or when certain statistical operations such as mean and variance are computed. We address this by employing an additional layer of protection - a non-linear and non-invertible feature transform - thereby providing a dual-layer security. Recent works have proposed methods to make deep learning computations more efficient in the FHE domain, like \cite{capride}, where the authors propose an encryption-friendly knowledge distillation method to create a framework for confidential and private decentralized training. In \cite{hetal}, the authors propose an encrypted transfer learning technique with efficient matrix multiplication and precise SoftMax approximation in the FHE domain. In \cite{fheIOT}, the authors propose to improve secure outsourcing computation in the IoT environment of medical devices using FHE, by optimizing the communication and computation burden in the IoT environment. In \cite{jain}, the authors develop a CNN using an FHE library to perform inference on encrypted data.    

\textbf{Hybrid Protection Techniques.}
A few papers in the literature have attempted to combine multiple privacy-preserving techniques to provide hybrid protection. In \cite{hybrid1}, the authors propose to combine cancelable biometric methods like bio hashing, MLP and index of maximum, along with FHE, for hybrid protection. In \cite{hybrid2}, the authors utilize homomorphic trans-ciphering to improve the security of FHE systems against offline decryption attacks, while \cite{hybrid3} uses a combination of bloom filters and homomorphic encryption for template protection.

\section{Proposed Approach}
First, we demonstrate that the embeddings computed from face images leak soft biometric information. For that purpose, we use two well-known face encoders, AdaFace \cite{adaface} and ArcFace \cite{arcFace}, examining the leakage of soft biometric attributes (age, gender, ethnicity) via their embeddings. While embedding protection methods (also known as template protection) have been proposed in the biometric literature, we show that several such non-linear and non-invertible feature transforms do not protect soft biometric attributes. Three such irreversible feature manifold hashes from the literature are used to demonstrate this. 

To address the pressing need for an effective privacy protection method that maintains a balance between privacy (of attributes) and utility (of face recognition), we introduce a novel robust shield centered around FHE-based computations on encrypted embeddings. Unlike conventional encryption methods, FHE supports computations on encrypted data, albeit with high computational overhead. To mitigate this, we incorporate embedding compression using Matryoshka Representation Learning (MRL), which retains core functionality in a reduced representation, thus enhancing computational efficiency within the encrypted domain. These compressed embeddings are subsequently encrypted via FHE, followed by the application of an irreversible feature manifold hash for additional security. The embedding hash ensures that the embedding will be protected by at least 1 layer of security in case of secret key leaks. In our study, we implement PolyProtect \cite{base} as the embedding hash and provide empirical evaluations to confirm the effectiveness of our framework.

\section{Experiments}
This section details experiments broadly based on the protection techniques described above. The goal in each of these experiments is to demonstrate the possibility of soft biometric leakage from face embeddings.

\subsection{Datasets}
We conducted our experiments using {\em CelebSet} \cite{CelebSet_Kaggle, celebset} and {\em Chicago Face Database (CFD)} \cite{chicagoFace}, with dataset statistics presented in Table \ref{CelebsetData} and Table \ref{CFDData}, respectively. As the CFD dataset does not provide age information, we employed the 'AgeRated' float values, rounding them to the nearest integer for consistency.

\begin{table*}[!ht]
\centering
\renewcommand{\arraystretch}{1.15}    
\resizebox{0.75\textwidth}{!}{
\begin{tabular}{cc|cccc|cccc}
\hline
\hline
\multicolumn{2}{c|}{\textbf{Gender}} & \multicolumn{4}{|c|}{\textbf{Age}} & \multicolumn{4}{|c}{\textbf{Ethnicity}} \\
\hline
\textbf{Males} & \textbf{Females} & \textbf{3-22} & \textbf{23-40} & \textbf{41-59} & \textbf{60-76} & \textbf{Hispanic} & \textbf{White} & \textbf{Black} & \textbf{Asian}\\
\hline
38080 & 34409 & 5279 & 43357 & 22781 & 1072 & 738 & 57873 & 13414 & 464\\ 
\hline
52.50\% & 47.50\% & 7.28\% & 59.82\% & 31.43\% & 1.47\% & 1.01\% & 79.85\% & 18.50\% & 0.64\%\\ 
\hline
\hline
\end{tabular}}
\caption{Statistics of the CelebSet dataset (80 Identities).}
\label{CelebsetData}
\end{table*}

\begin{table*}[!ht]

\centering
\renewcommand{\arraystretch}{1.15}    
\resizebox{0.75\textwidth}{!}{
\begin{tabular}{cc|cccc|cccc}
\hline
\hline
\multicolumn{2}{c|}{\textbf{Gender}} & \multicolumn{4}{|c|}{\textbf{Age (Rounded to the nearest integer)}} & \multicolumn{4}{|c}{\textbf{Ethnicity}} \\
\hline
\textbf{Males} & \textbf{Females} & \textbf{17-24} & \textbf{25-27} & \textbf{28-32} & \textbf{33-56} & \textbf{Asian} & \textbf{Black} & \textbf{Latino} & \textbf{White}\\ 
\hline
290 & 307  & 177 & 148 & 143 & 129  & 109 & 197 & 108 & 183\\ 
\hline
48.57\% & 51.42\% & 29.64\% & 24.79\% & 23.95\% & 29.61\% & 18.26\% & 32.99\% & 18.09\% & 30.65\%\\ 
\hline
\hline
\end{tabular}}
\caption{Statistics of the Chicago Face Database (597 Identities).}
\label{CFDData}
\end{table*}

\subsection{Irreversible Feature Manifold Hash techniques}
 In our study, we examine three template protection methods (embedding hash techniques) proposed in the literature, demonstrating that all of them still inadvertently leak soft biometric information. 1) PolyProtect (PP) \cite{base} transforms a face embedding to a more secure template, using a mapping based on multivariate polynomials parameterized by user-specific coefficients and exponents. 2) Negative Face Recognition (NFR) \cite{NFR} method produces two templates for each face, positive and negative, where the latter is privacy-protected. Face embeddings are enlarged to a size of 4096 and transformed to obtain a positive template. The negative template, which is the protected reference template, is obtained by randomly replacing each value in the positive template with any of the other values within the vector. 3) Minimum Information Units (MIU) \cite{MIU} partitions the face embedding into blocks of size $k$, and these blocks are shuffled to obtain a protected embedding. Table \ref{TP_leakage} shows the soft biometric leakage from embeddings protected by irreversible feature manifold hashes.

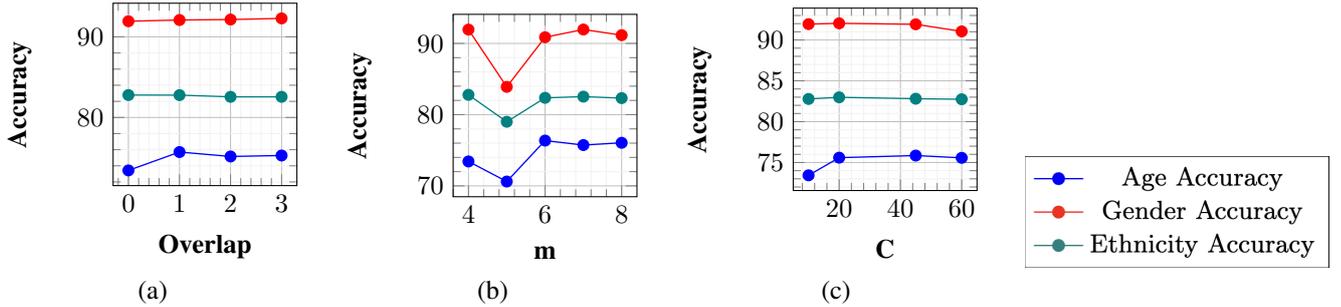
\begin{figure*}[!h]
\begin{tabular}{cccc}

        \centering
    \begin{tikzpicture}
    \begin{axis}[
      width=4cm,
      height=4cm,
      xlabel={Overlap},
      ylabel={Accuracy},
      font=\bfseries,
      legend pos=north east,
      legend style={font=\tiny, at={(0.98,0.80)}, anchor=north east},
      grid=both,
      grid style={line width=.1pt, draw=gray!10},
      major grid style={line width=.2pt,draw=gray!50},
      minor tick num=4,
    ]

    \addplot[mark=*, blue, line width = 0.5pt] table[x=Overlap, y=Age] {
      Overlap Age
      0 73.43
      1 75.71
      2 75.16
      3 75.28
    };

    \addplot[mark=*, red, line width = 0.5pt] table[x=Overlap, y=Gender] {
      Overlap Gender
      0 91.94
      1 92.09
      2 92.15
      3 92.29
    };

    \addplot[mark=*, teal, line width = 0.5pt] table[x=Overlap, y=Ethnicity] {
      Overlap Ethnicity
      0 82.78
      1 82.77
      2 82.56
      3 82.56
    };
    \end{axis}
  \end{tikzpicture}
  \label{overlap_pp}
&
        \begin{tikzpicture}
    \begin{axis}[
      width=4cm,
      height=4cm,
      xlabel={m},
      ylabel={Accuracy},
      font=\bfseries,
      legend pos=north east,
      legend style={font=\tiny, at={(0.98,0.98)}, anchor=north east},
      grid=both,
      grid style={line width=.1pt, draw=gray!10},
      major grid style={line width=.2pt,draw=gray!50},
      minor tick num=4,
    ]

    \addplot[mark=*, blue, line width = 0.5pt] table[x=m, y=Age] {
      m Age
      4 73.43
      5 70.61
      6 76.36
      7 75.73
      8 76.05
    };

    \addplot[mark=*, red, line width = 0.5pt] table[x=m, y=Gender] {
      m Gender
      4 91.94
      5 83.90
      6 90.87
      7 91.95
      8 91.16
    };

    \addplot[mark=*, teal, line width = 0.5pt] table[x=m, y=Ethnicity] {
      m Ethnicity
      4 82.78
      5 79.00
      6 82.36
      7 82.53
      8 82.32
    };
    \end{axis}
  \end{tikzpicture}
  \label{m_pp} &
        \begin{tikzpicture}
    \begin{axis}[
      width=4cm,
      height=4cm,
      xlabel={C},
      ylabel={Accuracy},
      font=\bfseries,
      legend pos=north east,
      legend style={font=\small, at={(0.98,0.98)}, anchor=north east},
      grid=both,
      grid style={line width=.1pt, draw=gray!10},
      major grid style={line width=.2pt,draw=gray!50},
      minor tick num=4,
    ]

    \addplot[mark=*, blue, line width = 0.5pt] table[x=C, y=Age] {
      C Age
      10 73.43
      20 75.59
      45 75.85
      60 75.57
    };

    \addplot[mark=*, red, line width = 0.5pt] table[x=C, y=Gender] {
      C Gender
      10 91.94
      20 92.04
      45 91.92
      60 91.05
    };

    \addplot[mark=*, teal, line width = 0.5pt] table[x=C, y=Ethnicity] {
      C Ethnicity
      10 82.78
      20 82.98
      45 82.81
      60 82.74
    };

    \end{axis}
  \end{tikzpicture}
  \label{C_pp}
        &
       \includegraphics[width=4cm]{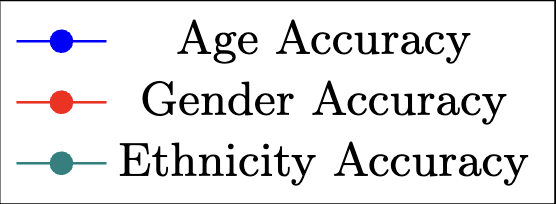} \\

      (a) & (b) & (c) & \\
    
        \end{tabular}
        \caption{Ablation Study with PolyProtect shows soft biometric leakage under different settings. (a) Overlap  (b) m - Length of polynomial coefficients/exponents (c) [-C,C] - Range of polynomial coefficients.}
        \label{fig:pp_ablation}
     
\end{figure*}

To validate our proposed multi-layer impregnable shield, we adopt PolyProtect (PP) \cite{base} across all experiments. PP operates with three primary parameters — `m', `C', and `overlap' — and following our ablation study in Fig. \ref{fig:pp_ablation}, we set $m=5$, $C=50$, and $overlap=0$ to maximize protection. We hypothesize that template protection schemes, which operate in the embedding space, can be applied to a variety of embeddings to ensure irreversibility and unlinkability \cite{base}.

\begin{table}

\centering
\renewcommand{\arraystretch}{1.15}    
\resizebox{0.45\textwidth}{!}{
\begin{tabular}{c|c|c|c|ccc}
\hline
\hline
\multirow{2}{*}{\textbf{Model}} & \multirow{2}{*}{\textbf{Dataset}} & \multirow{2}{*}{\shortstack{\textbf{Protection}\\ \textbf{technique}}} & \multirow{2}{*}{\textbf{Id (\%)}} & \multicolumn{3}{|c}{\textbf{Classification accuracy} ($\downarrow$)} \\
\cline{5-7}
& & &  & \cellcolor{lightblue}\textbf{Gen (\%)} & \textbf{Age (\%)} & \cellcolor{lightblue}\textbf{Ethn (\%)}\\
\hline
\multirow{6}{*}{ArcFace} & \multirow{3}{*}{CelebSet} & PP & 84.40 & \cellcolor{lightblue}85.60 & 66.90 & \cellcolor{lightblue}77.87 \\
& & MIU & 99.68 & \cellcolor{lightblue}84.06 & 66.70 & \cellcolor{lightblue}89.02\\
& & NFR & 99.69 & \cellcolor{lightblue}84.60 & 67.50 & \cellcolor{lightblue}66.13\\
\cline{2-7}
 & \multirow{3}{*}{CFD} & PP &85.04 & \cellcolor{lightblue}74.10 & 81.66 & \cellcolor{lightblue}58.33\\
 & & MIU &  84.41& \cellcolor{lightblue}71.90 & 83.34 & \cellcolor{lightblue}51.50\\
 & & NFR & 85.21 & \cellcolor{lightblue}75.00 & 87.50 & \cellcolor{lightblue}49.16\\
\hline

\multirow{6}{*}{AdaFace} & \multirow{3}{*}{CelebSet} & PP & 86.31 & \cellcolor{lightblue}93.93 & 71,35 & \cellcolor{lightblue}89.12\\
& & MIU & 91.37 & \cellcolor{lightblue}88.20 & 69.70 & \cellcolor{lightblue}79.40\\
& & NFR & 83.43 & \cellcolor{lightblue}88.68 & 70.13 & \cellcolor{lightblue}69.18\\
\cline{2-7}

 & \multirow{3}{*}{CFD} & PP &88.46 & \cellcolor{lightblue}80.80 & 82.50 & \cellcolor{lightblue}66.60\\
 & & MIU & 86.51& \cellcolor{lightblue}74.16 & 86.60 & \cellcolor{lightblue}52.50\\
 & & NFR &87.22 & \cellcolor{lightblue}83.33 & 87.50 & \cellcolor{lightblue}58.30\\
 \hline
\end{tabular}}
\caption{Soft biometric leakage in terms of attribute classification accuracy (\%) using various Template Protection schemes (CFD - Chicago Face Dataset; Gen - Gender; Ethn - Ethnicity; MIU - Minimum Information Units; NFR - Negative Face Recognition; PP - PolyProtect).}
\label{TP_leakage}
\end{table}


\pgfplotsset{compat=1.15}
\begin{figure}
  \centering
  \begin{tikzpicture}
    \begin{axis}[
      width=0.4\textwidth,
      height=0.3\textwidth,
      xlabel={Compression Dimension},
      ylabel={Classification Accuracy (\%)},
      font=\bfseries,
      legend pos=south east,
      legend style ={font=\small},
      xmin=8, xmax=256,
      ymin=50, ymax=100,
      xmode=log,
      xtick={8, 16, 32, 64, 128, 256},
      xticklabels={8, 16, 32, 64, 128, 256},
      grid=both,
      grid style={line width=.1pt, draw=gray!10},
      major grid style={line width=.2pt,draw=gray!50},
      minor tick num=4,
    ]

    \addplot[mark=*, blue, line width = 0.5pt] table[x=Compression, y=Identity] {
      Compression Identity
      8 58.12
      16 81.18
      32 88
      64 90.62
      128 91.25
      256 91.06
    };
    \addlegendentry{Identity}

    \addplot[mark=*, red, line width = 0.5pt] table[x=Compression, y=Age] {
      Compression Age
      8 66
      16 71.5
      32 73.5
      64 78.4
      128 78.13
      256 77.68
    };
    \addlegendentry{Age}

    \addplot[mark=*, teal, line width=0.5pt] table[x=Compression, y=Gender] {
      Compression Gender
      8 70.75
      16 84.81
      32 91.56
      64 95.81
      128 95.87
      256 97.06
    };
    \addlegendentry{Gender}

    \addplot[mark=*, purple, line width = 0.5pt] table[x=Compression, y=Ethnicity] {
      Compression Ethnicity
      8 82.93
      16 89.25
      32 94.69
      64 93.5
      128 96.5
      256 95.31
    };
    \addlegendentry{Ethnicity}

    \end{axis}
  \end{tikzpicture}
  \caption{Performance of attribute extraction (Age, Gender, Ethnicity), along with Rank-1 identification accuracy (Identity), after compression of embeddings using Matryoshka Representation Learning (MRL), as a function of different compression dimensions. The graph is based on an experiment performed using AdaFace on the CelebSet dataset.}
  \label{fig:mrl}
\end{figure}
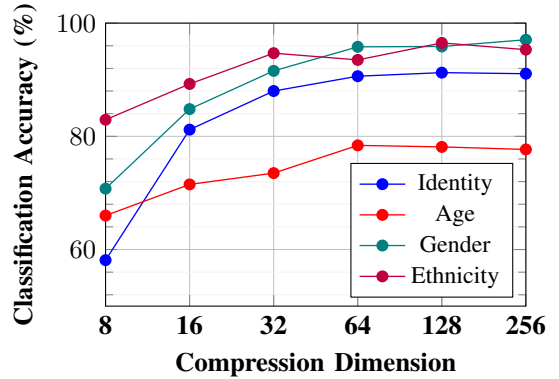


\pgfplotsset{compat=1.15}
\begin{figure}
  \centering
  \begin{tikzpicture}
    \begin{axis}[
      width=0.4\textwidth,
      height=0.35\textwidth,
      xlabel={Epsilon},
      ylabel={Classification Accuracy (\%)},
      font=\bfseries,
      legend pos=north west,
      legend style={
          font=\footnotesize,
          yshift=0.15cm 
      },
      xmode=log,
      grid=both,
      grid style={line width=.1pt, draw=gray!10},
      major grid style={line width=.2pt,draw=gray!50},
      minor tick num=4,
    ]

    \addplot[mark=*, blue, line width = 0.5pt] table[x=Epsilon, y=Identity] {
      Epsilon Identity
      0 98.89
      0.001 0.93
      0.01  1.2
      0.1   1.18
    0.25 1.80   
    0.5 1.88
    1 4.92
    2 9.65
    4 19.38
    8 38.81
    10 66.66
    };
    \addlegendentry{Identity}

    \addplot[mark=*, red, line width = 0.5pt] table[x=Epsilon, y=Age] {
      Epsilon Age
            0 89.95
            0.001 56.5
            0.01 56.81
            0.1 56.00
            0.25 56.01
            0.5 56.25
            1 56.25
            2 59.82
            4 68.00
            8 72.88
            10 77.25
    };
    \addlegendentry{Age}

    \addplot[mark=*, teal, line width=0.5pt] table[x=Epsilon, y=Gender] {
      Epsilon Gender
      0 96.66
      0.001 49.31
      0.01 52.75
      0.1 52.5
    0.25 50.81
    0.5 49.25
    1 52.68
    2 58.31
    4 70.88
    8 83.82
    10 89.68
    };
    \addlegendentry{Gender}

    \addplot[mark=*, purple, line width = 0.5pt] table[x=Epsilon, y=Ethnicity] {
      Epsilon Ethnicity
      0 96.96
      0.001 50.56
      0.01 49.00
      0.1 50.37
    0.25 51.44
    0.5 49.13
    1 50.75
    2 59.17
    4 66.76
    8 82.25
    10 87.07
    };
    \addlegendentry{Ethnicity}

    \end{axis}
  \end{tikzpicture}
  \caption{Classification accuracy of attributes (Age, Gender, Ethnicity) along with Rank-1 identification accuracy (Identity) of face embeddings protected by Differential Privacy as a function of Privacy Budget ($\epsilon$) and  Sensitivity ($\Delta f = 2$). The graph is based on an experiment performed using AdaFace on the CelebSet dataset.}
  \label{fig:dp}
\end{figure}
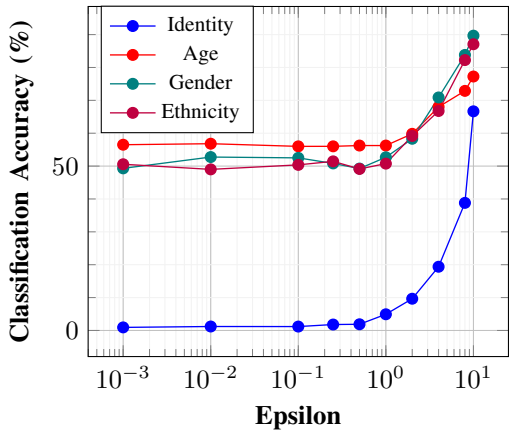


\subsection{Embedding Compression}
 We explore embedding compression through \textbf{Matryoshka Representation Learning (MRL}) \cite{matryoshka} as a technique to improve both privacy in the plaintext domain and processing speeds in the encrypted domain. Based on the ablation study in Fig. \ref{fig:mrl}, we can infer that a compressed embedding size of 64 retains the best performance across all attributes. We train an MRL model to compress embeddings while computing the loss based on identity. Subsequently, we train classifiers to extract soft biometric attributes and predict identity based on the compressed embeddings.

\subsection{Differential Privacy}
We investigate the efficacy of Differential Privacy in mitigating soft biometric leakage while maintaining biometric recognition accuracy. To introduce Differential Privacy into our embeddings, we employ the Laplacian mechanism governed by two parameters - sensitivity ($\Delta f$) and privacy budget ($\epsilon$) - as outlined in \cite{dp}\cite{dp_obfuscation_facial_images}. Since the maximum distance between two normalized embeddings is utmost 2, we set the sensitivity parameter to 2. From the ablation study in Fig. \ref{fig:dp}, we can infer that a privacy budget $\leq 10^{-1}$ provides maximum privacy. A noise vector of the same size as the embeddings is generated, composed of values drawn from the Laplacian distribution. Subsequently, we add this noise vector to the embeddings to form a DP-protected embedding. All downstream tasks are conducted with the DP-protected embeddings.

\subsection{Fully Homomorphic Encryption (FHE)}
\label{exp_fhe}
FHE is an encryption technique that allows computations on encrypted data. The permitted computations are restricted to additive and multiplicative operations only, thereby posing a significant challenge to conduct non-linear operations on ciphertexts. In this paper, we adapt all non-linear functions, such as activation functions, cosine similarity computations, and PolyProtect (PP) \cite{base}, in innovative ways to operate in the encrypted domain. We utilize the HEAAN library \cite{HEAAN} to perform all FHE experiments.

\textbf{PolyProtect in FHE.} PolyProtect (PP) is a multivariate polynomial function designed to operate in the plaintext domain. We adapt it to FHE by using Taylor series approximations and packing $m$ values of the embedding in a single ciphertext. Subsequently, we obtain a vector of ciphertexts, which are then accumulated with the appropriate ciphertext rotations.

\textbf{Primary task.} In our experiments, we focus on face identification as the primary task on the two datasets. We use the cosine distance metric to compute the match score between two {\em encrypted} (ciphertext) embeddings and a trained multi-layer perceptron to compute the match score between two {\em unencrypted} (plaintext) embeddings. First, embeddings from the training set undergo MRL+FHE+PP processing and are then stored as encrypted embeddings. Next, we compute the cosine distance between the encrypted test embeddings and the stored embeddings. Since cosine distance involves an inverse-square root calculation (making it a non-linear operation), we approximate this inverse square root function as an 8th-degree polynomial through polynomial regression to enable cosine distance computation in the encrypted domain. 

\textbf{Soft biometric attribute classification.} To evaluate the classification accuracy of soft biometric attributes from encrypted embeddings, we train SVM classifiers on the ASCII values of the encrypted embeddings. 

\subsection{Execution Times}

The execution times for all experiments are presented in Table \ref{execution_time}. Since HEAAN \cite{HEAAN} supports SIMD operations on ciphertext, we did not conduct separate experiments for each encoder. All plaintext domain experiments were performed using Python on a single A100 Nvidia GPU, while the FHE experiments were conducted with the HEAAN library \cite{HEAAN}, which runs on C++ and requires a Linux OS. It is important to note that the execution time for ``MRL + PP" experiment does not simply equal the sum of the execution times of MRL and PP experiments; this is because PP is applied to the compressed embedding in the ``MRL + PP" experiment. From Table \ref{execution_time}, we observe that compressing the embedding (MRL+FHE+PP) provides us with approximately 627\% speedup compared to the experiment without compression (FHE+PP). Although the execution times are considerably high, we assert that privacy should not be compromised for speed.

\begin{table}
    \centering
    \renewcommand{\arraystretch}{1.15}    
    \resizebox{0.45\textwidth}{!}{
    \begin{tabular}{c|c}
    \hline
    \hline
    \textbf{Experiment}  & \textbf{Execution Time (ms)} \\
    \hline
    MRL (Plaintext domain) & 0.5 \\ \hline
    PP (Plaintext domain) & 32 \\ \hline
    MRL + PP (Plaintext domain) & 13 \\ \hline
     FHE Encryption  & 85\\  \hline
     FHE Decryption & 45 \\ \hline
     FHE + PP & 5200 \\  \hline
     MRL (Plaintext domain) + FHE & 12 \\ \hline
     \shortstack{MRL (Plaintext domain) + FHE\\+ PP} & 715 \\ \hline
     \shortstack{Cosine Similarity of 512-sized\\embeddings (Identification)} & 9478 \\
     \hline
     \shortstack{Cosine Similarity of 64-sized\\embeddings (Identification)} & 1077\\
     \hline
     \hline
    \end{tabular}}
    \caption{We report the execution times of the proposed method — MRL+FHE+PP. Note that the execution time for MRL+PP does not simply equal the sum of the execution times of MRL and PP due to the embedding size compression in the MRL+PP approach.}
    \label{execution_time}
\end{table}

\begin{table*}[!ht]

\centering
\renewcommand{\arraystretch}{1.1}    
\resizebox{0.65\textwidth}{!}{
\begin{tabular}{c|c|c|c|ccc}
\hline
\hline
\multirow{2}{*}{\textbf{Model}} & \multirow{2}{*}{\textbf{Dataset}} & \multirow{2}{*}{\textbf{Protection technique}} & \multirow{2}{*}{\textbf{Id (\%)}} & \multicolumn{3}{|c}{\textbf{Classification accuracy} ($\downarrow$)} \\
\cline{5-7}
& & &  & \cellcolor{lightblue}\textbf{Gen (\%)} & \textbf{Age (\%)} & \cellcolor{lightblue}\textbf{Ethn (\%)}\\
\hline
\multirow{6}{*}{ArcFace} & \multirow{6}{*}{CelebSet} & None & 
99.42 & \cellcolor{lightblue} 98.12 & 87.68 & \cellcolor{lightblue}98.81\\ 
& & PP & 99.42 & \cellcolor{lightblue} 97.35 & 85.00 & \cellcolor{lightblue}98.06\\ 

& & MRL & 97.93 & \cellcolor{lightblue} 98.00 & 85.87 & \cellcolor{lightblue}97.38\\ 
& & DP & 1.28 & \cellcolor{lightblue} 49.49 & 59.23 & \cellcolor{lightblue}53.37\\ 

& & MRL+PP &97.00 & \cellcolor{lightblue} 96.32 & 87.32 & \cellcolor{lightblue}98.44\\ 

& & MRL+FHE & 97.83 & \cellcolor{lightblue} \textbf{52.22} & \textbf{6.12} & \cellcolor{lightblue}\textbf{8.01}\\ 

& & MRL+FHE+PP & 96.95 & \cellcolor{lightblue} \textbf{52.22} & \textbf{6.12} & \cellcolor{lightblue}8.03\\ 
\cline{2-7}

 & \multirow{6}{*}{CFD} & None & 85.06 & \cellcolor{lightblue} 95.25 & 94.40 & \cellcolor{lightblue}91.97\\
& & PP & 85.04 & \cellcolor{lightblue} 95.07 & 93.97 & \cellcolor{lightblue}91.62\\ 

& & MRL &84.42 & \cellcolor{lightblue} 92.02 & 93.98 & \cellcolor{lightblue}91.65\\ 
 & & DP & 0.20 & \cellcolor{lightblue} 51.23 & 58.65 & \cellcolor{lightblue}77.81\\ 

& & MRL+PP &84.31 & \cellcolor{lightblue} 90.00 & 87.90 & \cellcolor{lightblue}87.70\\ 

& & MRL+FHE &84.38 & \cellcolor{lightblue} 49.98 & 28.00 & \cellcolor{lightblue}24.01\\ 

& & MRL+FHE+PP & 84.28 & \cellcolor{lightblue} \textbf{49.50} & \textbf{27.82} & \cellcolor{lightblue}\textbf{24.00}\\

\hline

\multirow{6}{*}{AdaFace} & \multirow{6}{*}{CelebSet} & None & 99.41 &\cellcolor{lightblue} 96.93  &90.25 & \cellcolor{lightblue}96.31\\
& & PP &99.38 & \cellcolor{lightblue}95.75 & 90.18 & \cellcolor{lightblue}96.56\\

& & MRL & 97.95 & \cellcolor{lightblue}97.63 & 85.50 & \cellcolor{lightblue}98.94\\ 
& & DP & 1.32 & \cellcolor{lightblue}48.12 & 59.17 & \cellcolor{lightblue}54.66\\

& & MRL+PP & 97.59 & \cellcolor{lightblue}96.82 & 89.13 & \cellcolor{lightblue}98.25\\ 

& & MRL+FHE & 97.91 & \cellcolor{lightblue}\textbf{52.22} & 6.12 & \cellcolor{lightblue}8.02\\ 

& & MRL+FHE+PP & 97.55 & \cellcolor{lightblue}\textbf{52.22} & \textbf{6.11} & \cellcolor{lightblue}\textbf{8.01}\\ 
\cline{2-7}

 & \multirow{6}{*}{CFD} & None & 88.55 & \cellcolor{lightblue}94.97 & 95.02 & \cellcolor{lightblue}91.72\\
& & PP & 88.46 & \cellcolor{lightblue}94.07 & 94.62 & \cellcolor{lightblue}89.62\\
& & MRL & 88.33 & \cellcolor{lightblue}92.55 & 93.83 & \cellcolor{lightblue}87.43\\ 
& & DP & 0.17 & \cellcolor{lightblue}50.98 & 58.44 & \cellcolor{lightblue}80.88\\

& & MRL+PP & 88.29 & \cellcolor{lightblue}90.58 & 88.66 & \cellcolor{lightblue}85.58\\ 

& & MRL+FHE & 88.23 & \cellcolor{lightblue}49.98 & 27.97 & \cellcolor{lightblue}24.02\\ 

& & MRL+FHE+PP & 88.21 & \cellcolor{lightblue}\textbf{49.50} & \textbf{27.90} & \cellcolor{lightblue}\textbf{24.00}\\ 

\hline
\hline
\end{tabular}
}
\caption{Identification accuracy and soft biometric classification accuracy using different protection techniques (DP - Differential Privacy; FHE - Fully Homomorphic Encryption; MRL - Matryoksha Representation Learning; PP - PolyProtect). The highest score has been denoted in \textbf{bold}.}
\label{Accuracy}
\end{table*}

\section{Results}
To evaluate our framework for leakage, we have chosen two metrics: Suppression Rate (SR) and Privacy Gain (PG) \cite{leakageMetrics}\cite{leakageMetrics2}\cite{ leakageMetrics3}. Privacy Gain is defined as $\text{PG} = (1 - R_p) - (1-R_o)$, where, $R_o$ and $R_p$ represent the recognition performances on the original data and the privacy-enhanced data, respectively. A positive value of PG signifies enhanced data protection. The ideal value for PG is 1. Suppression Rate measures the difference in attribute prediction accuracy with and without privacy enhancement, $A_p$ and $A_o$, respectively: $\text{SR} = (A_o - A_p)/(A_o)$.

\begin{table*}

\centering
\renewcommand{\arraystretch}{1.1}    
\resizebox{0.85\textwidth}{!}{
\begin{tabular}{c|c|c|c|ccc|ccc}

\hline
\hline
\multirow{2}{*}{\textbf{Model}} & \multirow{2}{*}{\textbf{Dataset}} & \multirow{2}{*}{\textbf{Protection technique}} & \multirow{2}{*}{\textbf{Id(\%)}} & \multicolumn{3}{c|}{\textbf{Privacy Gain} ($\uparrow$)} & \multicolumn{3}{c}{\textbf{Suppression Rate} ($\uparrow$)}\\
\cline{5-10}
& & & & \cellcolor{lightblue} \textbf{Gender} & \textbf{Age} & \cellcolor{lightblue} \textbf{Ethn} & \textbf{Gender} & \cellcolor{lightblue} \textbf{Age} & \textbf{Ethn}\\
\hline
\multirow{6}{*}{ArcFace} & \multirow{6}{*}{CelebSet} &  PP & 99.42 & \cellcolor{lightblue}0.72 & 2.68 & \cellcolor{lightblue}0.75 & 0.0073 & \cellcolor{lightblue}0.0306 & 0.0075\\ 

& & MRL & 97.93 & \cellcolor{lightblue}0.12 & 1.81 & \cellcolor{lightblue}1.43 & 0.0012 & \cellcolor{lightblue}0.0206 & 0.0145\\

& & DP & 1.28 & \cellcolor{lightblue}\textbf{48.63}  & 28.45 & \cellcolor{lightblue}45.44 & \textbf{0.4956} & \cellcolor{lightblue}0.3244 & 0.4590 \\ 

& & MRL+PP &97.00 & \cellcolor{lightblue}1.80 & 0.36 & \cellcolor{lightblue}0.37 & 0.0183 & \cellcolor{lightblue}0.0041 & 0.0037\\ 

& & MRL+FHE & 97.83 & \cellcolor{lightblue}45.90 & \textbf{81.56} & \cellcolor{lightblue}\textbf{90.80} & 0.4678 & \cellcolor{lightblue}\textbf{0.9302} & \textbf{0.9189}\\ 

& & MRL+FHE+PP & 96.95 & \cellcolor{lightblue}45.90 &\textbf{81.56} & \cellcolor{lightblue}90.78 & 0.4678 & \cellcolor{lightblue}\textbf{0.9302} & 0.9187\\ 
\cline{2-10}

 & \multirow{6}{*}{CFD}  & PP & 85.04 & \cellcolor{lightblue}0.18 & 0.43 & \cellcolor{lightblue}0.35 & 0.0019 & \cellcolor{lightblue}0.0046 & 0.0038\\ 

& & MRL &84.42& \cellcolor{lightblue}3.23 & 0.42 & \cellcolor{lightblue}0.32 & 0.0339 & \cellcolor{lightblue}0.0044 & 0.0035\\ 

& & DP & 0.20 & \cellcolor{lightblue}44.02 & 35.75 & \cellcolor{lightblue}14.16  & 0.4600 &\cellcolor{lightblue}0.3700 &0.1500 \\

& & MRL+PP &84.31 & \cellcolor{lightblue}5.25 & 6.50 & \cellcolor{lightblue}4.27 & 0.0551 & \cellcolor{lightblue}0.0689 & 0.0464\\ 

& & MRL+FHE &84.38 & \cellcolor{lightblue}45.27 & 66.40 & \cellcolor{lightblue}67.96 & 0.4753 & \cellcolor{lightblue}0.7034 & 0.7389\\ 

& & MRL+FHE+PP & 84.28 & \cellcolor{lightblue}\textbf{45.75} & \textbf{66.58} & \cellcolor{lightblue}\textbf{67.97} & \textbf{0.4803} & \cellcolor{lightblue}\textbf{0.7053} & \textbf{0.7390}\\ 

\hline

\multirow{6}{*}{AdaFace} & \multirow{6}{*}{CelebSet} & PP &99.38 & \cellcolor{lightblue}1.18 & 0.07 & \cellcolor{lightblue}-0.25 & 0.0122 & \cellcolor{lightblue}0.0008 & -0.0026\\

& & MRL & 97.95 & \cellcolor{lightblue}-0.70 & 4.75 & \cellcolor{lightblue}-2.63 & -0.0072 & \cellcolor{lightblue}0.0526 & -0.0273\\ 

& & DP & 1.32 & \cellcolor{lightblue}\textbf{48.81}  & 31.08 & \cellcolor{lightblue}41.65  & \textbf{0.5000} & \cellcolor{lightblue}0.3400 & 0.4300  \\

& & MRL+PP & 97.59 & \cellcolor{lightblue}0.11 & 1.12 & \cellcolor{lightblue}-1.94 & 0.0011 & \cellcolor{lightblue}0.0124 & -0.0201\\ 

& & MRL+FHE & 97.91 & \cellcolor{lightblue}44.71 & 84.13 & \cellcolor{lightblue}88.29 & 0.4613 & \cellcolor{lightblue}0.9322 & 0.9167\\ 

& & MRL+FHE+PP & 97.55 & \cellcolor{lightblue}44.71 & \textbf{84.14} & \cellcolor{lightblue}\textbf{88.30} & 0.4613 & \cellcolor{lightblue}\textbf{0.9323} & \textbf{0.9168}\\ 
\cline{2-10}

 & \multirow{6}{*}{CFD} & PP & 88.46 & \cellcolor{lightblue}0.90 & 0.40 & \cellcolor{lightblue}2.10 & 0.0090 & \cellcolor{lightblue}0.0040 & 0.0220\\

& & MRL & 88.33 & \cellcolor{lightblue}2.42 & 1.19 & \cellcolor{lightblue}4.29 & 0.0250&  \cellcolor{lightblue}0.0120& 0.0460\\ 

& & DP & 0.17 & \cellcolor{lightblue}\textbf{43.99} & 36.58 & \cellcolor{lightblue}10.84  & \textbf{0.4632}& \cellcolor{lightblue}0.3840 &  0.1100\\

& & MRL+PP & 88.29 & \cellcolor{lightblue}4.39 & 6.36 & \cellcolor{lightblue}6.14 &  0.0460& \cellcolor{lightblue}0.0660 & 0.0670\\ 

& & MRL+FHE & 88.23 & \cellcolor{lightblue}39.99 & 57.05 & \cellcolor{lightblue}57.70 & 0.4444 & \cellcolor{lightblue}0.6710 & 0.7061\\ 

& & MRL+FHE+PP & 88.21 & \cellcolor{lightblue}40.47 & \textbf{57.12} & \cellcolor{lightblue}\textbf{57.72} & 0.4498 & \cellcolor{lightblue}\textbf{0.6718} & \textbf{0.7063}\\ 

\hline
\hline
\end{tabular}
}
\caption{Privacy Gain and Suppression Rate across different soft biometric attributes using different protection techniques (DP - Differential Privacy; FHE - Fully Homomorphic Encryption; MRL - Matryoksha Representation Learning; PP - PolyProtect). The highest score has been denoted in \textbf{bold}.}
\label{PrivacyGain}
\end{table*}

\begin{figure}[!ht]
    \begin{tabular}{cc}
             
                 \includegraphics[width=3.9cm]{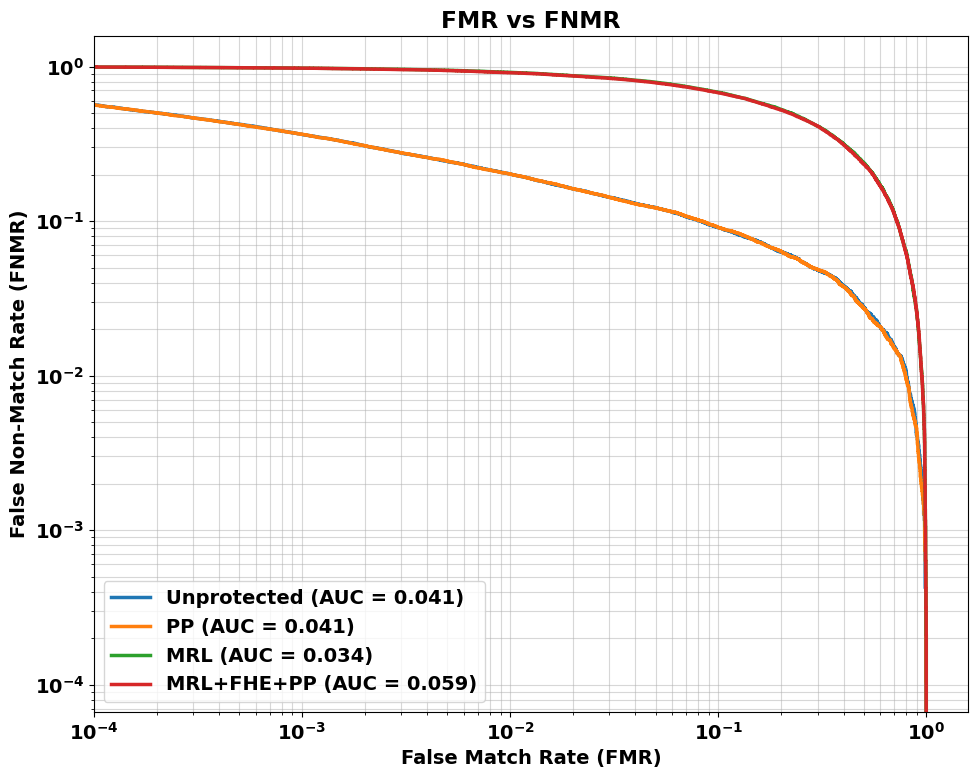}
                 
             &
                 \includegraphics[width=3.9cm]{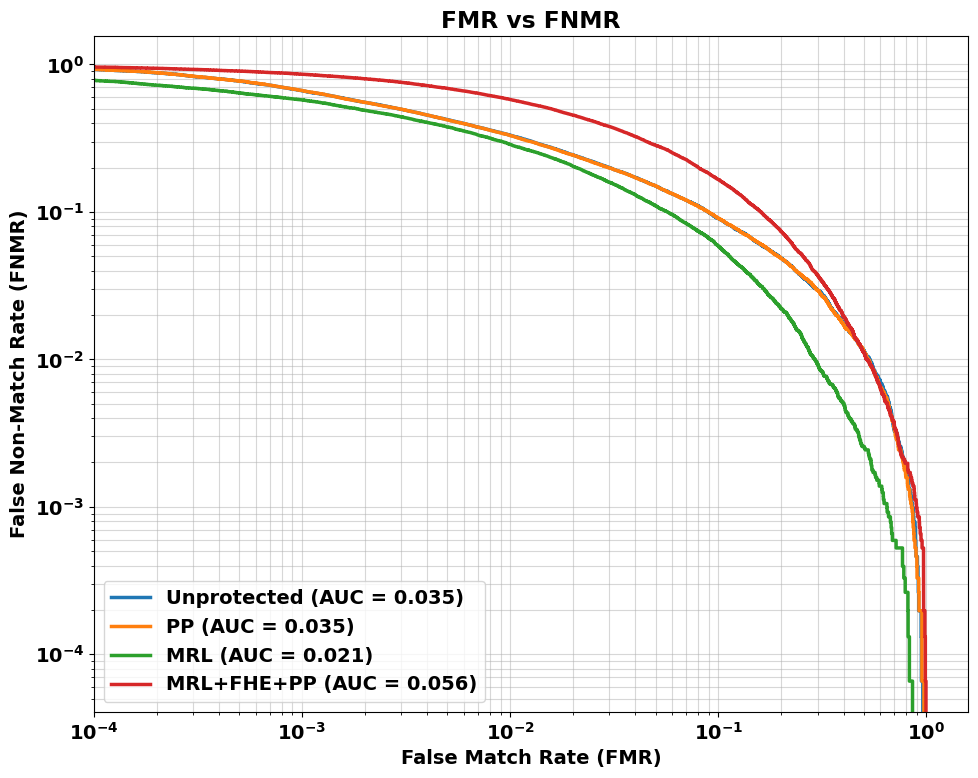}
                 \\
                 (a) & (b) \\ 
                 \includegraphics[width=3.9cm]{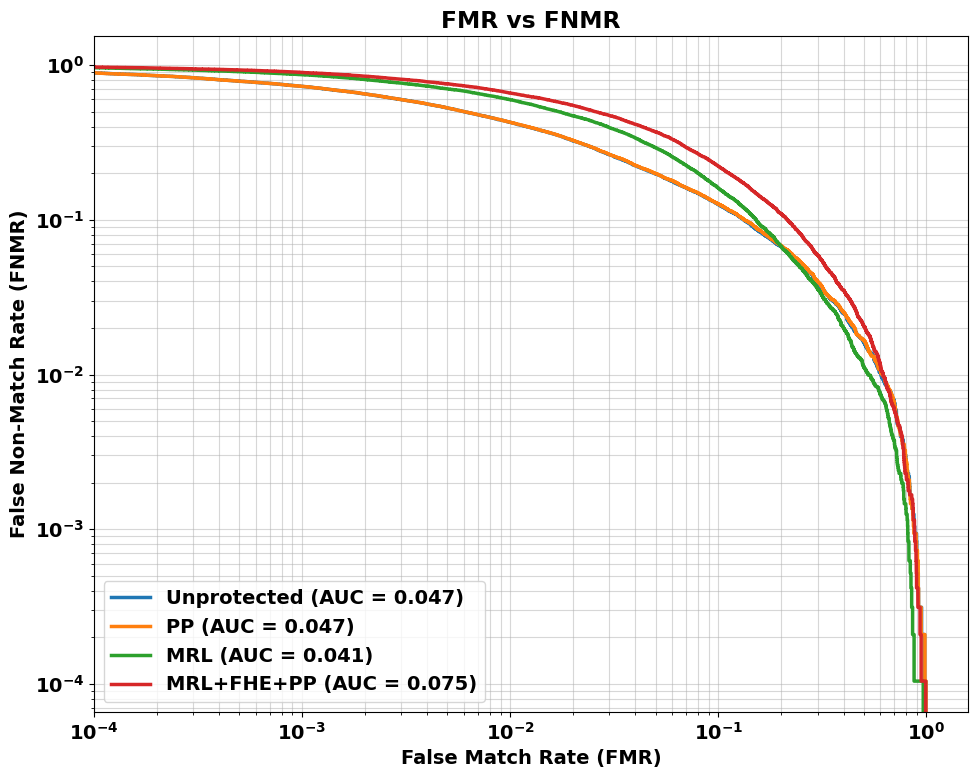}
                
             &
                 \includegraphics[width=3.9cm]{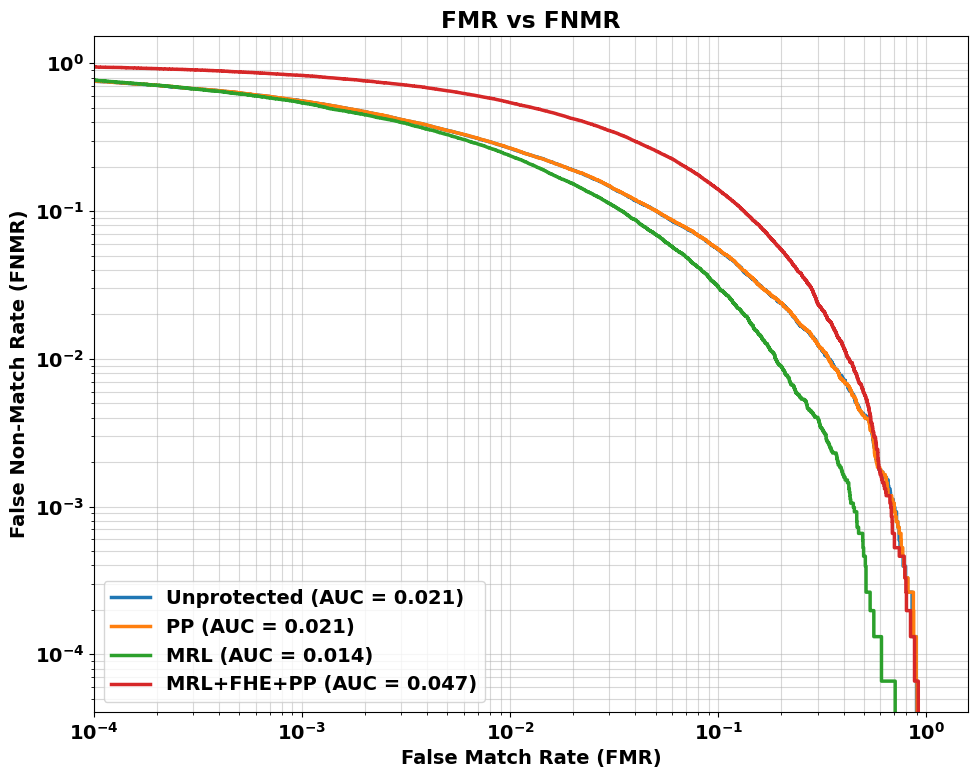}\\

     (c) & (d)\\
        \end{tabular}
        \caption{False Match Rate (FMR) vs False Non-Match Rate (FNMR) for identity verification performance using unprotected and privacy-protected embeddings - (a) AdaFace - CFD; (b) AdaFace - CelebSet; (c) ArcFace - CFD; (d) ArcFace - CelebSet. A smaller area-under-curve signifies better verification performance. PP curve overlaps the ``unprotected embeddings" curve, showing a negligible loss in identity verification performance.}
        \label{Fig:fmr_fmnr} 
\end{figure}

As evident in Tables \ref{Accuracy}, \ref{PrivacyGain}, Template Protection, Embedding Compression, and Differential Privacy, all leak a substantial amount of soft biometric information. On the other hand, our proposed approach of employing embedding compression through MRL, encryption using FHE, and template protection using PolyProtect demonstrates a high Privacy Gain. This shows that it prevents soft biometric leakage with minimal loss in face recognition accuracy. Fig \ref{Fig:fmr_fmnr} plots the FMR vs FMNR curve for face verification , where a smaller Area-Under-Curve indicates better performance. We can also observe that our approach reduced the classification accuracies of soft biometric attributes to the level of random chance in both datasets. We could also achieve an almost ideal Privacy Gain in certain scenarios, but we believe this could be due to the imbalanced nature of our datasets.

From Table \ref{auc_scores}, we observe soft biometric leakage in all scenarios except when embeddings are encrypted. In cases where embeddings are protected through homomorphic encryption, the AUC score remains below 0.55 across all instances, indicating that the classification performance is reduced to near-random levels. Since gender and ethnicity classification problems are multi-class, we have used macro-averaging to compute the AUC score.

\begin{table}
    \centering
    \renewcommand{\arraystretch}{1.1}    
    \resizebox{0.48\textwidth}{!}{
        \begin{tabular}
        {c|c|c|ccc}
            \hline
            \hline
             \textbf{Model} & \textbf{Dataset} & \textbf{Protection Technique} & \textbf{Age ($\downarrow$)} & \cellcolor{lightblue}\textbf{Gen ($\downarrow$)} & \textbf{Ethn ($\downarrow$)} \\
             \hline

             \multirow{6}{*}{ArcFace} & \multirow{3}{*}{CelebSet} & None & 0.88 & \cellcolor{lightblue}0.99 & 0.99 \\
             & & PP & 0.98 & \cellcolor{lightblue}0.99 & 0.99 \\
             & & MRL & 0.88 & \cellcolor{lightblue}0.99 & 0.99 \\
             & & DP & 0.52 & \cellcolor{lightblue}\textbf{0.47} & \textbf{0.50} \\
             & & MRL+PP & 0.97 & \cellcolor{lightblue}0.99 & 0.98 \\
             & & MRL+FHE & \textbf{0.46} & \cellcolor{lightblue}0.50 & 0.52\\
             & & MRL+FHE+PP & \textbf{0.46} & \cellcolor{lightblue}0.50 & 0.52 \\
             \cline{2-6}
              & \multirow{3}{*}{CFD} & None & 0.98 & \cellcolor{lightblue}0.99 & 0.98 \\
             & & PP & 0.99 & \cellcolor{lightblue}0.99 & 0.98 \\
             & & MRL & 0.98 & \cellcolor{lightblue}0.99 & 0.99 \\
             & & DP & \textbf{0.50} & \cellcolor{lightblue}\textbf{0.49} & 0.51 \\          
             & & MRL+PP & 0.99 & \cellcolor{lightblue}0.99 & 0.98 \\
             & & MRL+FHE & \textbf{0.50} & \cellcolor{lightblue}0.50 & 0.49 \\
             & & MRL+FHE+PP & 0.51 & \cellcolor{lightblue}\textbf{0.49} & \textbf{0.47}\\
             
             \hline
             
             \multirow{6}{*}{AdaFace} & \multirow{3}{*}{CelebSet} & None &0.84 & \cellcolor{lightblue} 0.99 & 0.99\\
             & & PP & 0.88 & \cellcolor{lightblue} 0.99 & 0.99\\
             & & MRL & 0.88 & \cellcolor{lightblue} 0.99 & 0.99\\
             & & DP & \textbf{0.49} & \cellcolor{lightblue} \textbf{0.48} & 0.51\\
             & & MRL+PP & 0.90 & \cellcolor{lightblue} 0.99 & 0.99\\
             & & MRL+FHE & 0.50 & \cellcolor{lightblue} 0.52 & \textbf{0.49}\\
             & & MRL+FHE+PP &\textbf{0.49} & \cellcolor{lightblue} 0.51 & 0.50\\
             \cline{2-6}
              & \multirow{3}{*}{CFD} & None & 0.92 & \cellcolor{lightblue} 0.99 & 0.97 \\
             & & PP & 0.98 & \cellcolor{lightblue} 0.99 & 0.99\\
             & & MRL & 0.94 & \cellcolor{lightblue} 0.99 & 0.98\\
             & & DP & 0.49 & \cellcolor{lightblue} 0.52 & 0.50\\
             & & MRL+FHE & \textbf{0.48} & \cellcolor{lightblue} 0.51 & \textbf{0.47}\\
             & & MRL+FHE+PP & \textbf{0.48} & \cellcolor{lightblue}\textbf{0.49} & 0.50 \\
             \hline
             \hline
        \end{tabular}}
    \caption{Area-Under-Curve (AUC) scores for soft biometric classification using various protection techniques (DP - Differential Privacy; FHE - Fully Homomorphic Encryption; MRL - Matryoksha Representation Learning; PP - PolyProtect). The lowest score has been denoted in \textbf{bold}}
    \label{auc_scores}
\end{table}

\section{Analysis}
\textbf{Why is the combination of FHE+PP not considered?}
MRL's contribution to privacy gain is negligible, but, it contributes significantly in improving processing time. From Table \ref{execution_time}, we can observe a 627\% speedup of our framework with embedding compression. The execution times for all FHE combinations are shown in Table \ref{execution_time}.

\textbf{How does our framework evaluate against DP?} DP provides maximum protection to our embeddings, as shown in Table \ref{Accuracy} - the accuracy values indicate randomness. However, our application requires us to conserve the primary functionality of embeddings (i.e., face identification or verification) while protecting soft biometric attributes. With DP, the embeddings lose their primary functionality. On the other hand, our proposed framework retains the primary task of the embeddings.

\textbf{Why is the order MRL+FHE+PP important?}
MRL is applied first to reduce the dimensionality of the embedding, thereby improving the computational efficiency of FHE. We choose to apply PP on FHE-encrypted embedding for privacy reasons. In cases where multiple PP embeddings and user parameters are leaked to the attacker, the likelihood of inversion increases \cite{base}. On the other hand, when the embeddings are encrypted before the application of PP, it would not compromise privacy.

\begin{figure}
    \centering
    \resizebox{0.45\textwidth}{!}{ 
        \begin{tabular}{cc}
        \footnotesize
            \begin{tikzpicture}
                \begin{axis}[
                    ybar,
                    symbolic x coords={Identity, Age, Gender, Ethnicity},
                    xtick=data,
                    x=1.1cm, 
                    ymin=0,
                    ylabel={Accuracy (\%)},
                    bar width=0.1cm, 
                    enlarge x limits=0.15, 
                    nodes near coords,
                    nodes near coords align={vertical},
                    every node near coord/.append style={rotate=90, anchor=west}
                ]
                    \addplot coordinates {(Identity, 99.42 ) (Age, 87.68) (Gender, 98.12) (Ethnicity, 98.81)};
                    \addplot coordinates {(Identity, 99.42 ) (Age, 85.00) (Gender, 97.35) (Ethnicity, 98.06)};
                    \addplot coordinates {(Identity, 1.28 ) (Age, 59.23) (Gender, 49.49) (Ethnicity, 53.37)};
                    \addplot coordinates {(Identity, 96.95) (Age, 6.12) (Gender, 52.22) (Ethnicity, 8.03)};
                \end{axis}
            \end{tikzpicture} 
            &
            \footnotesize
            \begin{tikzpicture}
                \begin{axis}[
                    ybar,
                    symbolic x coords={Identity, Age, Gender, Ethnicity},
                    xtick=data,
                    x=1.1cm,
                    ymin=0,
                    bar width=0.1cm,
                    enlarge x limits=0.15,
                    nodes near coords,
                    nodes near coords align={vertical},
                    every node near coord/.append style={rotate=90, anchor=west}
                ]
                    \addplot coordinates {(Identity, 85.06 ) (Age, 94.40) (Gender, 95.25) (Ethnicity, 91.97)};
                    \addplot coordinates {(Identity, 85.04 ) (Age, 93.97) (Gender, 95.07) (Ethnicity, 91.62)};
                    \addplot coordinates {(Identity, 0.20 ) (Age, 58.65) (Gender, 51.23) (Ethnicity, 77.81)};
                    \addplot coordinates {(Identity, 84.28) (Age, 27.82) (Gender, 49.50) (Ethnicity, 24.00)};
                \end{axis}
            \end{tikzpicture}\\
            \textbf{(i)} & \textbf{(ii)}\\
            \footnotesize
            \begin{tikzpicture}
                \begin{axis}[
                    ybar,
                    symbolic x coords={Identity, Age, Gender, Ethnicity},
                    xtick=data,
                    x=1.1cm,
                    ymin=0,
                    ylabel={Accuracy (\%)},
                    bar width=0.1cm,
                    enlarge x limits=0.15,
                    nodes near coords,
                    nodes near coords align={vertical},
                    every node near coord/.append style={rotate=90, anchor=west}
                ]
                    \addplot coordinates {(Identity, 99.41 ) (Age,90.25) (Gender, 96.93) (Ethnicity, 96.31)};
                    \addplot coordinates {(Identity, 99.38 ) (Age,90.18) (Gender, 95.75) (Ethnicity, 96.56)};
                    \addplot coordinates {(Identity,1.32 ) (Age, 66.31) (Gender, 70.87) (Ethnicity, 85.37)};
                    \addplot coordinates {(Identity, 97.55) (Age, 6.11) (Gender, 52.22) (Ethnicity, 8.01)};
                \end{axis}
            \end{tikzpicture}
            &
            \footnotesize
            \begin{tikzpicture}
                \begin{axis}[
                    ybar,
                    symbolic x coords={Identity, Age, Gender, Ethnicity},
                    xtick=data,
                    x=1.1cm,
                    ymin=0,
                    bar width=0.1cm,
                    enlarge x limits=0.15,
                    nodes near coords,
                    nodes near coords align={vertical},
                    every node near coord/.append style={rotate=90, anchor=west}
                ]
                    \addplot coordinates {(Identity, 88.55 ) (Age,95.02) (Gender, 94.97) (Ethnicity, 91.72)};
                    \addplot coordinates {(Identity, 88.46 ) (Age,94.62) (Gender, 94.07) (Ethnicity, 89.62)};
                    \addplot coordinates {(Identity, 0.17 ) (Age, 50.98) (Gender, 58.44) (Ethnicity, 80.88)};
                    \addplot coordinates {(Identity, 88.21) (Age, 27.90) (Gender, 49.50) (Ethnicity, 24.00)};
                \end{axis}
            \end{tikzpicture}\\
            \textbf{(iii)} & \textbf{(iv)} \\
        \end{tabular}
    }

    \resizebox{0.48\textwidth}{!}{
    \footnotesize
    \centering
        \begin{tikzpicture}
            \node[draw, fill=blue!30, minimum size=10pt] at (0.5,0) {};
            \node[anchor=west] at (0.8,0) {No Protection};
            
            \node[draw, fill=red!30, minimum size=10pt] at (3.5,0) {};
            \node[anchor=west] at (3.8,0) {PP};
            
            \node[draw, fill=brown!30, minimum size=10pt] at (5,0) {};
            \node[anchor=west] at (5.3,0) {DP};
            
            \node[draw, fill=gray!90, minimum size=10pt] at (6.5,0) {};
            \node[anchor=west] at (6.8,0) {MRL+FHE+PP};
        \end{tikzpicture}
    }

    \caption{Identification accuracy (Identity) and attribute classification accuracy (Age, Gender, Ethnicity) across different privacy techniques: (i) ArcFace on CelebSet; (ii) ArcFace on CFD; (iii) AdaFace on CelebSet; (iv) AdaFace on CFD.}
    \label{fig:accuracy_barplot}
\end{figure}
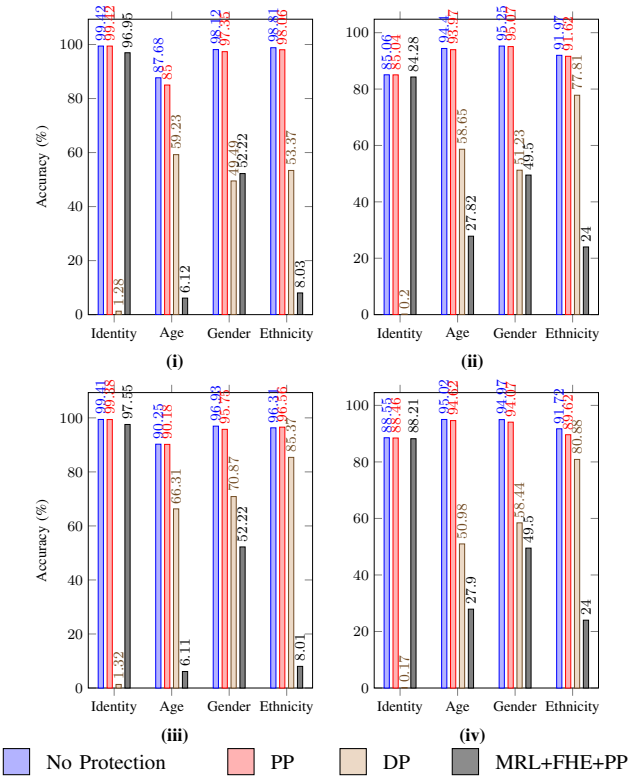

\section{Conclusion}
In this study, we investigated various protection techniques to secure face embeddings from potential attacks and privacy breaches. Our experimental results indicate that no single approach — whether Template Protection, Differential Privacy, or Fully Homomorphic Encryption — is sufficient to independently safeguard embeddings effectively. To address this, we propose a robust, multi-layered protection strategy combining (1) Embedding Compression, (2) Fully Homomorphic Encryption, and (3) Irreversible Feature Manifold Hash applied within the FHE space. This integrated approach achieves optimal embedding security by shielding soft biometric attributes while maintaining high biometric recognition accuracy, as validated through comprehensive experiments across two datasets and two distinct face encoders. By leveraging FHE’s theoretical guarantees, this strategy ensures that only authorized individuals with the correct secret key can access computation results. Moreover, even when the secret key is compromised, the irreversible feature transform serves as an additional safeguard to prevent full exposure of the embeddings.

\section{Ethical Impact Statement}
The overall approach in our research is focused on protecting leakage of privacy in facial analytics. Publicly available face datasets (CelebSet \cite{CelebSet_Kaggle}\cite{celebset} and Chicago Face Database \cite{chicagoFace}) have been used for reporting accuracy and performance evaluation. We do not show any of the dataset images / personal data in the paper. Our research does not involve data collection from human subjects. The contributions do not include the creation of a new dataset. To the best of our knowledge, we do not see any potential negative implications on the society from our paper.

{\small
\balance
\bibliographystyle{ieee}
\bibliography{FG2025}
}

\end{document}